\documentclass[journal=jacsat,manuscript=article]{achemso}
\usepackage[version=3]{mhchem} 
\usepackage{mathrsfs}[10pt]
\usepackage[latin1]{inputenc}
\usepackage{amsmath}
\usepackage{amsfonts}
\usepackage{amssymb}
\usepackage{algorithm}
\usepackage{float}
\usepackage{graphicx}
\usepackage{url}
\usepackage{multirow}
\usepackage{setspace}
\usepackage{caption}
\usepackage{color}

\def\tpr{\text{Pr}}
\def\setG{\mathcal{G}}
\def\setNG{\bar{\mathcal{G}}}

\def\citep{\cite}
\newcommand{\tabincell}[2]{\begin{tabular}{@{}#1@{}}#2\end{tabular}}

\author{Chao Yang}
\affiliation[The Hong Kong University of Science and
Technology]{Laboratory for Bioinformatics and Computational Biology,
Department of Electronic and Computer Engineering, The Hong Kong
University of Science and Technology, Hong Kong, China}
\author{Zengyou He}
\affiliation[Dalian University of Technology]{School of Software, Dalian University of Technology, Dalian, China}
\author{Weichuan Yu}
\affiliation[The Hong Kong University of Science and
Technology]{Laboratory for Bioinformatics and Computational Biology,
Department of Electronic and Computer Engineering, The Hong Kong
University of Science and Technology, Hong Kong, China}
\email{eeyu@ust.hk}

\title[Combinatorial Perspective]{A Combinatorial Perspective of the Protein Inference Problem}

\abbreviations{IR,NMR,UV}

\begin{document}
\begin{abstract}
In a shotgun proteomics experiment, proteins are the most biologically meaningful output. The success of proteomics studies depends on the ability to accurately and efficiently identify proteins. Many methods have been proposed to facilitate the identification of proteins from the results of peptide identification. However, the relationship between protein identification and peptide identification has not been thoroughly explained before.

In this paper, we are devoted to a combinatorial perspective of the protein inference problem. We employ combinatorial mathematics to calculate the conditional protein probabilities (Protein probability means the probability that a protein is correctly identified) under three assumptions, which lead to a lower bound, an upper bound and an empirical estimation of protein probabilities, respectively. The combinatorial perspective enables us to obtain a closed-form formulation for protein inference.

Based on our model, we study the impact of unique peptides and degenerate peptides on protein probabilities. Here, degenerate peptides are peptides shared by at least two proteins. Meanwhile, we also study the relationship of our model with other methods such as ProteinProphet. A probability confidence interval can be calculated and used together with probability to filter the protein identification result. Our method achieves competitive results with ProteinProphet in a more efficient manner in the experiment based on two datasets of standard protein mixtures and two datasets of real samples.

We name our program ProteinInfer. Its Java source code is available at:

\noindent\url{http://bioinformatics.ust.hk/proteininfer}
\end{abstract}

\section{Introduction}
Proteomics is developed to study the gene and cellular function directly at the protein level \citep{msproteomics}. In proteomics, mass spectrometry has been a primary tool in conducting high-throughput experiments. In a typical shotgun proteomic experiment, proteins are digested into peptides by enzymes and analyzed by a mass spectrometry to generate single stage mass spectra \citep{msproteomics,link1999direct,gygil1999quantitative}. Some peptides are fragmented into smaller ions and analyzed to produce tandem mass spectra. Identifying peptides from tandem mass spectra leads to the development of peptide identification methods \citep{sequest,perkins1999probability,craig2004tandem,geer2004open}. Protein inference is to derive proteins from peptide identification. Conducting protein identification in an accurate and high-throughput manner is a primary goal of proteomics \citep{rappsilber2002does,nesvizhskii2005interpretation}.

Quantitative measurement of protein identification confidence has been a major concern in developing new protein inference models. The calculation of protein probability has become popular as it provides nice properties in terms of distinction and accuracy:
\begin{itemize}
\item Distinction: Protein probability is a quantitative measurement of protein identification confidence that different proteins are distinguishable based on their probabilities.
\item Accuracy: By assigning each protein with a probability, we can have a statistical interpretation of the protein identification result. Thus, the protein identification result can be more reliable.
\end{itemize}

To date, many statistical models for protein probability calculation have been proposed \citep{huang2012protein}. The readers may refer to a recent review for detailed description on these methods \citep{serang2012areview}. Depending on the strategy to obtain protein probabilities, they can be grouped into the following two categories:
\begin{itemize}
\item Probability Models: Methods in this category partitioned the degenerate peptide probability among corresponding proteins. Degenerate peptides are peptides shared by at least two proteins. Then, they calculated the probability of a protein as the probability that at least one of its peptides was present. The partition weights were iteratively updated in an EM-like algorithm \citep{nesvizhskii2003statistical,searle2010scaffold,price2007ebp,feng2007probability}.
\item Bayesian Methods: Methods in this group modeled the process of mass spectrometry in a generative way by using the rigorous Bayesian framework \citep{shen2008hierarchical,li2010nested,yong2008bayesian,gerster2010protein}. The Bayesian models are complicated in their formulation. To obtain the solution, computational methods such as Markov chain Monte Carlo (MCMC) are essential. More importantly, methods such as MSBayesPro need extra information. The performance depends on the reliability of extra information and the methods may not be applicable under different conditions. Fido provides a way to calculate the marginal protein probability based on the Bayesian formula. However, the computational complexity of the method is exponential with respect to the number of distinct peptides.\cite{serang2010efficient}
\end{itemize}

In this paper, we provide a combinatorial perspective of the protein inference problem to calculate protein probabilities as well as to understand the probability partition procedure. Our contributions can be described in the following aspects:
\begin{itemize}
\item By computing the marginal protein probability, we have a concise protein inference model with an analytical solution, whose computational complexity is linear to the number of distinct peptide.
\item We deduce a lower bound and an upper bound of protein probability. Two bounds define a probability confidence interval, which is used as an alternative factor other than protein probability in filtering the protein identification result.
\item The impacts of unique peptides and degenerate peptides are studied mathematically based on our model. We also discuss some promising ways to further improve the distinction of protein inference.
\item We discuss the connection of our method with other methods such as greedy methods and ProteinProphet \cite{zhang2007proteomic, zengyouhe2011partial,nesvizhskii2003statistical}.
\end{itemize}

The rest of our paper is organized as follows: Section 2 describes the details of our method; Section 3 presents the experimental results; Section 4 discusses and concludes the paper.

\section{Method}
\begin{figure}[H]
\caption{The data structure of a protein identification problem. We need to estimate protein probabilities given peptides probabilities and peptide-protein mapping. In the figure, peptide 1 and peptide 2 are degenerate peptides whereas peptide 3 and peptide M are unique peptides.\label{fig:proteinidinput}}
\centerline{\includegraphics[width=\linewidth]{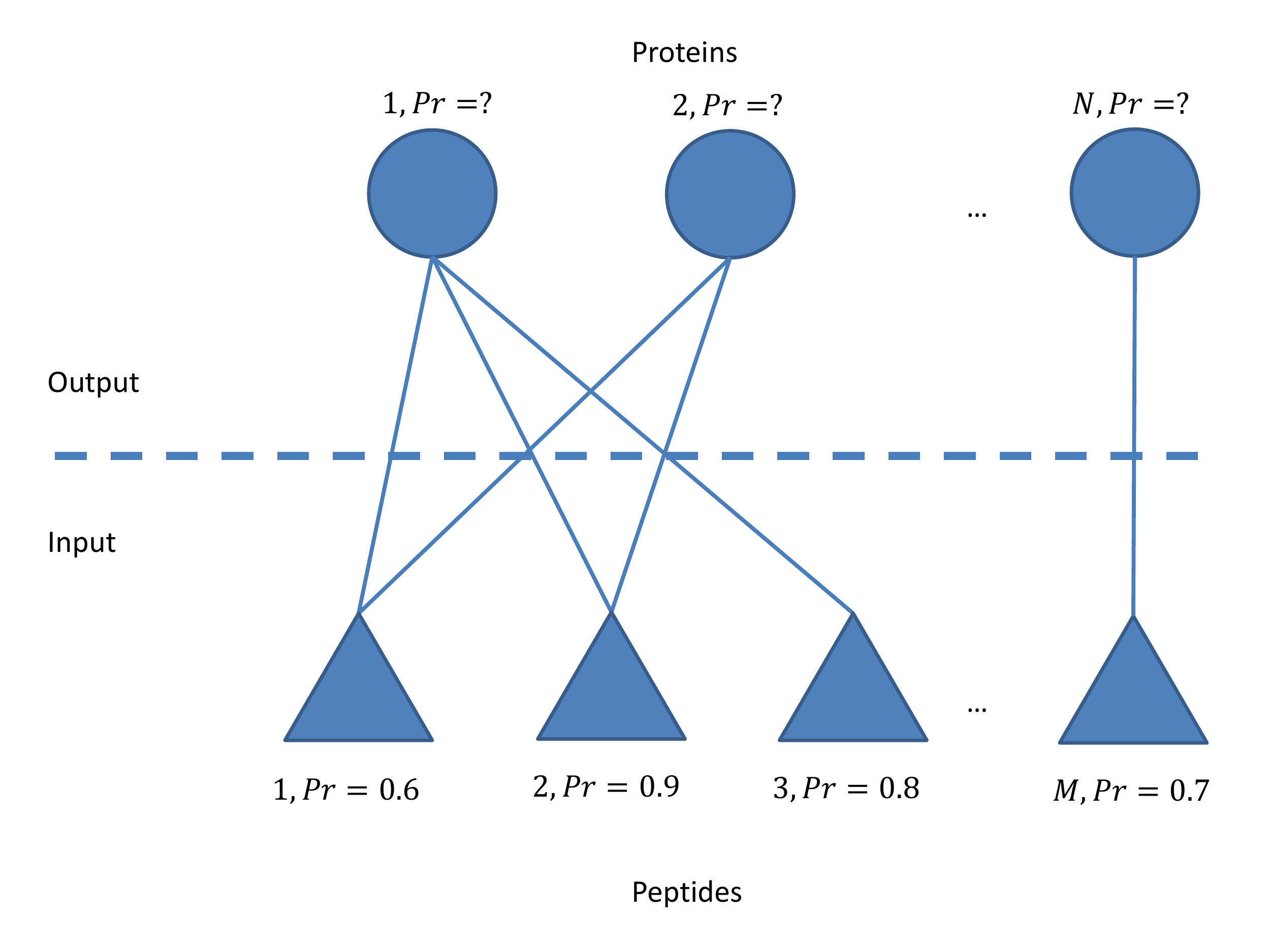}}
\end{figure}
Peptides from the same protein are assumed to contribute independently to presence of the protein. The assumption relates to Naive Bayes, in which features of an object contribute independently to the object. From the viewpoint of statistics, the independence assumption not only simplifies the modeling procedure, but also leads to stable results when information such as the inner structure of data is insufficient. The presence and the absence of peptides from different proteins are also independent.

Figure \ref{fig:proteinidinput} shows the data structure of the protein inference problem. In a protein identification result, there are many candidate proteins. For simplicity, we would like to focus on one protein to keep our notations concise.

Suppose a protein has $M$ peptides. We use $\tpr(y=1)$ and $\tpr(y=0)$ to denote the present probability and absent probability of the protein, respectively. For peptide $i$, $\tpr(x_i=1)$ and $\tpr(x_i=0)$ represent its present and absent probabilities, respectively. Let $n_i$ be the number of proteins which share peptide $i$. When $n_i\geq 2$, peptide $i$ is a degenerate peptide; otherwise, it is a unique peptide. We use $\setG=\{i|x_i=1\}$ and $\setNG=\{i|x_i=0\}$ to denote the set of present peptide indices and the set of absent peptide indices, respectively. For each pair of $\setG$ and $\setNG$, we have $\setG\cup\setNG=\{1,2,...,M\}$. When $i\in\setG$, it means that peptide $i$ is present (i.e. $x_i=1$). Similarly, $i\in\setNG$ means the absence of peptide $i$ (i.e. $x_i=0$). Denote the set $\{x_i|i\in{\setG}\}$ and the set $\{x_i|i\in{\setNG}\}$ as $X_{\setG}$ and $X_{\setNG}$, respectively. The conditional probability of the absence of the protein given peptides is denoted as $\tpr(y=0|X_{\setG},X_{\setNG})$. Let $\tpr(X_{\setG})$ and $\tpr(X_{\setNG})$ be $\prod_{i\in\setG}{\tpr(x_i)}$ and $\prod_{i\in\setNG}{\tpr(x_i)}$, respectively.

Each peptide of a protein is assumed to contribute independently to the protein. According to the basic probability theorem, the probability of a protein being absent is calculated as:
\begin{equation}
\label{eq:03}
\begin{split}
&\tpr(y=0)=\sum_{x_1\in\{0,1\}}\sum_{x_2\in\{0,1\}}...\sum_{x_M\in\{0,1\}}{\tpr(y=0|x_1,x_2,...,x_M)\prod_{i=1}^M{\tpr(x_i)}}\\
&=\sum_{X_{\setG},X_{\setNG}}{\tpr(y=0|X_{\setG},X_{\setNG})Pr(X_{\setG})Pr(X_{\setNG})}.
\end{split}
\end{equation}

To calculate the protein probability given by equation (1), we need to calculate the conditional probability $\tpr(y=0|X_{\setG},X_{\setNG})$. Directly computing the protein probability based on equation (1), we need $2^M$ operations. In the following sections, we will provide a way to calculate the protein probability with the number of operations that is linear with respect to $M$.

When $\setG$ is empty:
\begin{equation}
\label{eq:allabsent}
\begin{split}
\tpr(y=0|X_{\setNG})=\tpr(y=0|x_1=0,x_2=0,...,x_M=0)=1.
\end{split}
\end{equation}
When $\setG$ is not empty, different assumptions lead to different results. In the following sections, we calculate the conditional protein probabilities based on three different assumptions. These three assumptions lead to an upper bound, a lower bound and an empirical estimation of protein probability, respectively.

\subsection{Conditional Probability Based on A Loose Assumption}
A loose assumption supposes that, when peptide $i$ is detected, all corresponding proteins are present. In other words, this peptide is contributed by all corresponding proteins. This assumption is related to the one-hit rule used in protein inference \citep{gupta2009false}.

When $\setG$ is not empty, the corresponding protein must be present. Thus, the conditional absent probability is:
\begin{equation}
\label{eq:maximalocc}
\tpr(y=0|X_{\setG},X_{\setNG})=0.
\end{equation}
The absent probability is calculated as:
\begin{equation}
\label{eq:maximalprob}
\begin{split}
\tpr(y=0)=&\prod_{i=1}^M{\tpr(x_i=0)} + \sum_{X_{\setG},X_{\setNG},\setG\neq \varnothing}{\tpr(y=0|X_{\setG},X_{\setNG})Pr(X_{\setG})Pr(X_{\setNG})}\\
=&\prod_{i=1}^M{\tpr(x_i=0)}\\
=&\prod_{i=1}^M{(1-\tpr(x_i=1))}.
\end{split}
\end{equation}

The loose assumption leads to an upper bound of the protein probability:
\begin{equation}
\label{eq:upper}
\tpr_U(y=1)=1-\prod_{i=1}^M{(1-\tpr(x_i=1))}.
\end{equation}

\subsection{Conditional Probability Based on A Strict Assumption}
A strict assumption supposes that, if a peptide is detected, it only comes from one corresponding protein containing the peptide. Or equivalently, the peptide occurs for just once. This concept is mostly related to greedy protein inference methods, which iteratively select a protein that explains most of remaining peptides and removes peptides that have been explained \citep{zhang2007proteomic, zengyouhe2011partial}.

When $\setG$ is not empty, the total number of ways to explain observed peptides (i.e. peptides with corresponding indices in $\setG$) is given by:
\begin{equation}
\label{eq:totalsingle}
N_t = \prod_{i\in\setG}{\binom{n_i}{1}}=\prod_{i\in\setG}{n_i}.
\end{equation}
Here, $n_i$ is the number of proteins containing peptide $i$. When the protein is absent, the number of times that peptide $i$ is shared is decreased by 1. Then, the total number of ways to explain the observed peptides is:
\begin{equation}
\label{eq:absentsingle}
N_a = \prod_{i\in\setG}{\binom{n_i-1}{1}}=\prod_{i\in\setG}{(n_i-1)}.
\end{equation}
The conditional probability is given by:
\begin{equation}
\label{eq:probsingle}
\tpr(y=0|X_{\setG},X_{\setNG})=\frac{N_a}{N_t}=\frac{\prod_{i\in\setG}{(n_i-1)}}{\prod_{i\in\setG}{n_i}}.
\end{equation}

We can consider two special cases to get some intuitions from equation (8).

Case 1: If there is any unique peptide (i.e. $n_i=1$), then $\tpr(y=0|X_{\setG},X_{\setNG})=0$. This indicates that the protein must be present if the corresponding unique peptide is observed.

Case 2: Suppose $\setG=\{1\}$ and $\setNG=\{2,3,...,M\}$. If the first peptide is shared by an infinity number of proteins (i.e. $n_1\to\infty$), then we have:
\begin{equation}
\label{eq:specialsingle}
\tpr(y=0|X_{\setG},X_{\setNG})=\lim_{n_1\to\infty}\frac{n_1-1}{n_1}=1.
\end{equation}
Equation (9) indicates that, if the first peptide is shared by too many proteins, we cannot determine exactly the corresponding protein. The probability of the absence of the protein is 1.

When the conditional probability (8) is applied, we have:
\begin{equation}
\label{eq:equalsingle}
\begin{split}
&\tpr(y=0)=\sum_{X_{\setG},X_{\setNG}}{\tpr(y=0|X_{\setG},X_{\setNG})\tpr(X_{\setG})\tpr(X_{\setNG})}\\
&=\prod_{i=1}^M(1-\frac{1}{n_i}\tpr(x_i=1)).
\end{split}
\end{equation}
Details are available in our supplementary document.

This assumption is very strict and leads to a lower bound of protein probability:
\begin{equation}
\label{eq:upper}
\tpr_L(y=1)=1-\prod_{i=1}^M(1-\frac{1}{n_i}\tpr(x_i=1)).
\end{equation}

\subsection{Conditional Probability Based on A Mild Assumption}
A mild assumption supposes that, all proteins containing the peptide may generate the peptide. The presence of a peptide is contributed by either one or multiple proteins.

When $\setG$ is not empty, the total number of ways to explain observed peptides (i.e. peptides with corresponding indices in $\setG$) is given by:
\begin{equation}
\label{eq:totalway}
N_t = \prod_{i\in\setG}{\left[\sum_{k=1}^{n_i}{\binom{n_i}{k}}\right]}=\prod_{i\in\setG}{(2^{n_i}-1)}.
\end{equation}
Here, $n_i$ is the number of proteins that share peptide $i$. When the protein is absent, the number of times that peptide $i$ is shared is $n_i-1$. Thus, the number of ways to explain the peptides above is given by:
\begin{equation}
\label{eq:absent}
N_a = \prod_{i\in\setG}{(2^{(n_i-1)}-1)}.
\end{equation}
Then, we have:
\begin{equation}
\label{eq:prob}
\tpr(y=0|X_{\setG},X_{\setNG})=\frac{N_a}{N_t}=\frac{\prod_{i\in\setG}{(2^{(n_i-1)}-1)}}{\prod_{i\in\setG}{(2^{n_i}-1)}}.
\end{equation}

Similarly, let us consider two special cases to get some insights from equation (14).

Case 1: If there is any unique peptide (i.e. $n_i=1$), then $\tpr(y=0|X_{\setG},X_{\setNG}))=0$. The corresponding protein must be present to explain this unique peptide.

Case 2: Suppose $\setG=\{1\}$ and $\setNG=\{2,3,...,M\}$. If the first peptide is shared by an infinity number of number of proteins (i.e. $n_1\to\infty$), then we have:
\begin{equation}
\label{eq:special}
\tpr(y=0|X_{\setG},X_{\setNG})=\lim_{n_1\to\infty}\frac{2^{(n_1-1)}-1}{2^{n_1}-1}=\frac{1}{2}.
\end{equation}
Equation (15) has a meaningful interpretation. If peptide 1 is shared by an infinity number of proteins, determining the presence of a corresponding protein is like random guessing. The probabilities of the presence and absence of this protein are therefore both 0.5.

The strict assumption is exclusive. If a degenerate peptide has already been explained, other proteins are not considered. In contrast, the mild assumption is inclusive. Explaining a degenerate peptide with one protein will not affect the presence of other proteins containing this degenerate peptide. Thus, we obtain different results in equation (9) and equation (15).

The absent probability based on the mild assumption reads:
\begin{equation}
\label{eq:equal}
\begin{split}
&\tpr(y=0)=\sum_{X_{\setG},X_{\setNG}}{\tpr(y=0|X_{\setG},X_{\setNG})\tpr(X_{\setG})\tpr(X_{\setNG})}\\
&=\prod_{i=1}^M(1-\frac{2^{n_i}}{2(2^{n_i}-1)}\tpr(x_i=1)).
\end{split}
\end{equation}
The proof can be found in the supplementary document.

The assumption leads to an empirical estimation of protein probability:
\begin{equation}
\label{eq:empirical}
\tpr_E(y=1)=1-\prod_{i=1}^M(1-\frac{2^{n_i}}{2(2^{n_i}-1)}\tpr(x_i=1)).
\end{equation}

\subsection{Marginal Protein Probability}
The relationship of the three protein probabilities based on the three assumptions above is:
\begin{equation}
\label{eq:trueprob}
\begin{split}
&\tpr_L(y=1)=1-\prod_{i=1}^M(1-\frac{1}{n_i}\tpr(x_i=1))\\
&\leq \tpr_E(y=1)=1-\prod_{i=1}^M(1-\frac{2^{n_i}}{2(2^{n_i}-1)}\tpr(x_i=1))\\
&\leq \tpr_U(y=1)=1-\prod_{i=1}^{M}{(1-\tpr(x_i=1))}.
\end{split}
\end{equation}
Here, $n_i\geq 1$ is the number of times that peptide $i$ is shared. Readers can refer to our supplementary document for the proof. The closed-form inequality (18) can be used to calculate the lower bound, the empirical estimation and the upper bound of protein probability efficiently. The total numbers of operations to calculate $\tpr_L$, $\tpr_E$ and $\tpr_U$ are linear with respect to the number of distinct peptide. The equality is achieved when all peptides of the protein are unique peptides. The empirical protein probability $\tpr_E(y=1)$ is used as a major factor for measuring the protein identification confidence. The difference between the upper bound and the lower bound is:
\begin{equation}
\label{eq:probdiff}
\tpr_D(y=1)=\tpr_U(y=1)-\tpr_L(y=1).
\end{equation}
The difference $\tpr_D(y=1)$ can be used to measure the confidence of the estimation. The smaller the value of $\tpr_D(y=1)$, the higher the confidence. When all peptides are unique, the probability estimation has no ambiguity and $\tpr_D(y=1)=0$. In this case, the confidence is the highest.

Different from exiting protein probability estimation methods, we have quantitative measurements of protein identification confidence from different aspects. This makes it possible to achieve superior distinction in the protein identification result.

\subsection{Unique Peptides and Degenerate Peptides}
Unique peptides play central roles in protein identification. We are more confident at the identification when more peptides from this protein are unique. According to our calculation (18), the highest confidence of a protein is achieved when its peptides are all unique.

Degenerate peptides can increase the empirical protein probability $\tpr_E$. However, the degree of the increase depends on how many times the degenerate peptides are shared by other proteins. To see this, let us consider a protein with unique peptide indices being $\{1,2,...,M-1\}$ and a degenerate peptide $M$. According to equation (16), we have:
\begin{equation}
\label{eq:05}
\begin{split}
&\tpr(y=0)=\prod_{i=1}^M(1-\frac{2^{n_i}}{2(2^{n_i}-1)}\tpr(x_i=1))\\
&=\prod_{i=1}^{M-1}(1-\tpr(x_i=1))(1-\frac{2^{n_M}}{2(2^{n_M}-1)}\tpr(x_M=1))\\
&\leq \prod_{i=1}^{M-1}{\tpr(x_i=0)}.
\end{split}
\end{equation}
Thus, degenerate peptide $M$ increases $\tpr_E$:
\begin{equation}
\label{eq:06}
\tpr_E(y=1)=1-\tpr(y=0)\geq 1-\prod_{i=1}^{M-1}{\tpr(x_i=0)}.
\end{equation}
The absent probability in equation (20) is monotonically increasing with respect to $n_M$, which results in the monotonically decreasing in $\tpr_E$.

Degenerate peptide $M$ introduces ambiguity in protein probability estimation. The confidence interval is given by:
\begin{equation}
\label{eq:diffdegenerate}
\begin{split}
&\tpr_D(y=1)=\tpr_U(y=1)-\tpr_L(y=1)\\
&=\prod_{i=1}^{M-1}{\tpr(x_i=0)}\left[\frac{n_M-1}{n_M}\tpr(x_M=1)\right].
\end{split}
\end{equation}
We can see that the ambiguity $\tpr_D$ increases when $n_M$ increases.

In conclusion, a degenerate peptide improves the empirical protein probability $\tpr_E$ and introduces ambiguity in protein probability calculation (i.e. $\tpr_D\neq 0$). As $n_M$ increases, the increase in $\tpr_E$ becomes smaller and the confidence interval $\tpr_D$ becomes larger.

\subsection{The Relationship with ProteinProphet}
In our model, the loose assumption and the strict assumption relate to the one-hit rule and greedy methods, respectively. In this section, we discuss the relationship of our method with ProteinProphet.

ProteinProphet calculates the protein probability as the probability that at least one of its peptides is present. When processing a degenerate peptide, ProteinProphet apportions this peptide among all proteins which share it. The protein with higher probability is assigned with more weight. The probability of the absence of the protein is then formulated as:
\begin{equation}
\label{eq:proteinprophet}
\tpr(y=0)=\prod_{i=1}^M{(1-w_i\tpr(x_i=1))}.
\end{equation}
Here, $w_i$ is the weight assigned to peptide $i$ of the protein. The assumption that the protein is present when at least one of its peptide is present is questionable when degenerate peptides are detected. Thus, ProteinProphet intuitively partitions the probability of degenerate peptide $i$ according to weight $w_i$. Then, a dummy peptide with probability $w_i\tpr(x_i)$ is assumed to be unique. Dummy peptides together with original unique peptides are all unique. Under this situation, the assumption that the protein is present if any peptide is detected is correct. This is because the protein must be present to explain its unique dummy or original peptides.

By comparing the ProteinProphet protein scoring function with our model (18), we can see that they are very related. In the initial stage, ProteinProphet evenly apportions degenerate peptides among all corresponding proteins (i.e. $w_i=\frac{1}{n_i}$). That is exactly the lower bound $\tpr_L$ in our model. At last, low confident proteins tend to occupy less weight whereas high confident proteins tend to occupy more weight. Generally for high confident proteins, we have $\frac{1}{n_i}\leq w_i\leq 1$. Thus, probabilities of these proteins estimated by ProteinProphet will be within the bound of our model. Although ProteinProphet is not strictly originated from the basic probability theorem, its formulation coincides with our model (18). This explains the popularity and good performance of ProteinProphet in real applications.

\section{Results}
In this section, we first describe our experimental settings such as datasets and tools used in the experiments. Then, we illustrate the unique peptide probability adjustment, which is a preprocessing step in protein inference. Next, we explain the reporting format of our protein identification result. Finally, we present and discuss the experimental results.

\subsection{Experimental Settings}
The evaluation of our method is conducted on four public available datasets: ISB, Sigma49, Human and Yeast. The ISB dataset was generated from a standard protein mixture which contains 18 proteins \citep{dataset}. The sample was analyzed on a Waters/Micromass Q-TOF using an electrospray source. The Sigma49 dataset was acquired by analyzing 49 standard proteins on a Thermo LTQ instrument. The Human dataset was obtained by analyzing human blood serum samples with Thermo LTQ. The Yeast dataset was obtained by analyzing cell lysate on both LCQ and ORBI mass spectrometers from wild-type yeast grown in rich medium. The information of each dataset is shown in Table \ref{tab:dataset}.

\begin{table}[H]
\caption{Names and URLs of Data Files.\label{tab:dataset}}
\centering
\begin{tabular}{lll}
\hline
Dataset & File Name&URL\\
\hline
\footnotesize ISB & $\footnotesize \text{QT20051230\_S\_18mix\_04.mzXML}$&\tabincell{c}{http://regis-web.systemsbiology.net\\/PublicDatasets/}\\
\footnotesize Sigma49&$\footnotesize \text{Lane/060121Yrasprg051025\-ct5.RAW}$&\tabincell{c}{https://proteomecommons.org/\\dataset.jsp?i=71610}\\
\footnotesize Human &\footnotesize  $\text{PAe000330/021505\_LTQ10401\_1\_2.mzXML}$&\tabincell{c}{http://www.peptideatlas.org/\\repository/}\\
\footnotesize Yeast & \footnotesize \tabincell{c}{$\text{YPD\_ORBI/}$061220.zl.mudpit0.1.1/\\raw/000.RAW}&\tabincell{c}{http://aug.csres.utexas.edu/msnet/}\\
\hline
\end{tabular}
\end{table}

When analyzing the ISB dataset and the Sigma49 dataset, we use the curve of false positives versus true positives to evaluate the performance. The ground truth of the ISB dataset and the Sigma49 dataset contains 18 and 49 proteins, respectively. A protein identification is a true positive if it is from ground truth. Otherwise, the protein is a false positive. Given the same number of false positives, more true positives mean a better performance. When analyzing the Human dataset and the Yeast dataset, we prefer the curve of decoys versus targets because the ground truth is not known in advance. Given the same number of decoys, the more the targets, the better the performance. We also plot the curves of decoys versus targets for the ISB dataset and the Sigma49 dataset.

The database we use is a target-decoy concatenated protein database, which contains 1048840 proteins. The decoys are obtained by reversing protein sequences of UniProtKB/Swiss-Prot (Release 2011\_01). In our experiments, X!Tandem (Version 2010.10.01.1) is employed to identify peptides from each dataset. In database search, the parameters ``fragment monoisotopic mass error'', ``parent monoisotopic mass error plus'' and ``parent monoisotopic mass error minus'' are set to be 0.4Da, 2Da and 4Da, respectively. The number of missed cleavages permitted is 1. Then, PeptideProphet and iProphet embedded in TPP (Version v4.5 RAPTURE rev 2, Build 201202031108 (MinGW)) are used to estimate peptide probabilities \citep{keller2002empirical,shteynberg2011iprophet,pedrioli2010trans}. At last, ProteinProphet and our method are applied to estimate protein probabilities. The performance of ProteinProphet is compared to that of our method.

\subsection{Adjust Unique Peptide Probabilities}
Protein inference models take the peptide identification results as input. If the peptide probability estimation is perfect, peptide probability adjustment is not essential. However, inferior peptide probabilities always exist.

Unique peptides are important for protein identification. A confident misidentified unique peptide (i.e. $\tpr(x=1)=0.99$) will result in a high confident protein identification with a high $\tpr_E$ and a low $\tpr_D$. For example, if a tandem mass spectrum is matched to a decoy peptide, the peptide is very likely to be unique. The unique high confident decoy peptide will lead the decoy protein to be identified with a high confidence. This motivates the procedure of unique peptide adjustment as a preprocessing step of our method.

Suppose a protein has $m$ unique peptides. The adjusted unique peptide probability can be calculated as:
\begin{equation}
\label{eq:pepadjust}
\tpr(x_i=1|m)=\frac{\tpr(m|x_i=1)\tpr(x_i=1)}{\tpr(m|x_i=1)\tpr(x_i=1)+\tpr(m|x_i=0)\tpr(x_i=0)}.
\end{equation}
Here, peptide $i$ is a unique peptide; $\tpr(x_i=1)$ is the probability that the unique peptide is true. The terms $\tpr(m|x_i=1)$ and $\tpr(m|x_i=0)$ describe the probabilities of observing $m$ unique peptides of the protein when the unique peptide $i$ is a true and a false identification, respectively. We model $\tpr(m|x_i=1)$ and $\tpr(m|x_i=0)$ as Poisson distributions with different expected number of unique peptides (i.e. $\lambda_1$ and $\lambda_2$):
\begin{equation}
\label{eq:poisson}
\begin{split}
\tpr(m|x_i=1)&=\frac{\lambda_1^me^{-\lambda_1}}{m!}\\
\tpr(m|x_i=0)&=\frac{\lambda_2^me^{-\lambda_2}}{m!}
\end{split}.
\end{equation}
Generally, a true unique peptide tends to have more sibling unique peptides than a false unique peptide on average. Thus, we have $\lambda_1>\lambda_2$. In our program, these two parameters can be manually specified. Alternatively, these two parameters can be obtained empirically.

Suppose there are $N$ candidate proteins and the number of unique peptides of protein $j (j\in\{1,2,...,N\})$ is $m_j$. The empirical value of the expected value $\lambda_1$ is estimated as:
\begin{equation}
\label{eq:lambda1}
\lambda_1=\frac{\sum_{j=1}^N{I(m_j\geq2)m_j}}{\sum_{j=1}^N{I(m_j\geq2)}}.
\end{equation}
Here, $I(\cdot)$ is an indicator function with value being either 0 or 1.

Empirically, $\lambda_2$ can be 1. It is common to observe that false proteins such as decoy proteins to have a single unique peptide.

The adjusted unique peptide probability $\tpr(x_i=1|m)$ is used as the probability of peptide $i$ in our model (18).

\subsection{The Protein Identification Result}
There are three different kinds of relationships between two proteins:
\begin{itemize}
\item Indistinguishable: If two proteins contain exactly the same set of identified peptides, they are indistinguishable. Indistinguishable proteins can be treated as a group.
\item Subset: Identified peptides of a protein form a peptide set. If the peptide set of a protein is the subset of the other, the former protein is a subset protein of the latter.
\item Differentiable: Two proteins are differentiable if they both contain peptides those are not from the other.
\end{itemize}

In the literature, subset proteins are generally discarded. However, it is not reasonable to regard all subset proteins as being absent from a statistical point of view. Thus, we also calculate the protein probability of subset proteins and organize our result in two separate files as shown in Table \ref{tab:01}. In the table, proteins $1$ and $3$ are indistinguishable proteins and protein $2$ is a subset protein of protein $1$.

\begin{table}[H]
\centering
\caption{Subset and Non-subset Proteins.\label{tab:01}}
{\fontsize{12pt}{10pt}\selectfont
\begin{tabular}{ccccccc}
\hline
\multicolumn{7}{c}{Non-subset Proteins}\\
\hline
Index&Protein&$\tpr_E$&$\tpr_L$&$\tpr_U$&$\tpr_D$&Other Proteins\\
1&$1$&$\tpr_E(y_1)$&$\tpr_L(y_1)$&$\tpr_U(y_1)$&$\tpr_D(y_1)$&$3$\\
2&$4$&$\tpr_E(y_4)$&$\tpr_L(y_4)$&$\tpr_U(y_4)$&$\tpr_D(y_4)$&-\\
\multicolumn{7}{c}{......}\\
\hline
\multicolumn{7}{c}{Subset Proteins}\\
\hline
Index&Protein&$\tpr_E$&$\tpr_L$&$\tpr_U$&$\tpr_D$&Subset of Protein\\
1&$2$&$\tpr_E(y_{2})$&$\tpr_L(y_{2})$&$\tpr_U(y_{2})$&$\tpr_D(y_{2})$&$1$\\
\multicolumn{7}{c}{......}\\
\hline
\end{tabular}
}
\end{table}

The empirical probability $\tpr_E$ and the bound $\tpr_D$ quantitatively describe the confidence of a protein. Generally, a high $\tpr_E$ and a low $\tpr_D$ mean a confident protein. The identification result is mainly sorted by the empirical protein probability $\tpr_E$. In case two proteins have the same $\tpr_E$, the order is then determined by $\tpr_D$.

One purpose of protein identification is to select proteins to explain observed peptides. Subset proteins do not increase the peptide explanation power. In general, we can perform downstream analysis by only using non-subset proteins. However, according to the probability theorem, the probability of a subset protein is not necessarily smaller than that of a non-subset protein. The data explanation power and protein probability are totally two different kinds of things. Without any prior knowledge, choosing proteins from data explanation's viewpoint is safe. However, in some cases, we may consider high confident subset proteins according to the prior knowledge we have. For instance, the sample contains homogeneous proteins and it is possible that subset proteins are present.

\begin{figure}[H]
\caption{An example of the protein identification result. In the figure, there are two proteins and three peptides with corresponding probabilities all being 0.9. Protein 2 is a subset protein of protein 1. The probability $\tpr_E$ and the bound $\tpr_D$ are two quantitative measurements of the confidence of a protein. The higher the value of $\tpr_E$ and the smaller the value of $\tpr_D$, the more confident the protein. Protein 1 is a confident protein with a high empirical probability $\tpr_E=0.984$ and a tight bound $\tpr_D=0.029$. For protein 2, there are two peptides present. Without any prior knowledge, we cannot determine the presence of protein 2 mathematically. From the data explanation's aspect, we can report protein 1 only. Protein 2 can be considered if the protein coverage is a concern and homogeneous proteins are known to be present.\label{fig:example}}
\centerline{\includegraphics[width=\linewidth]{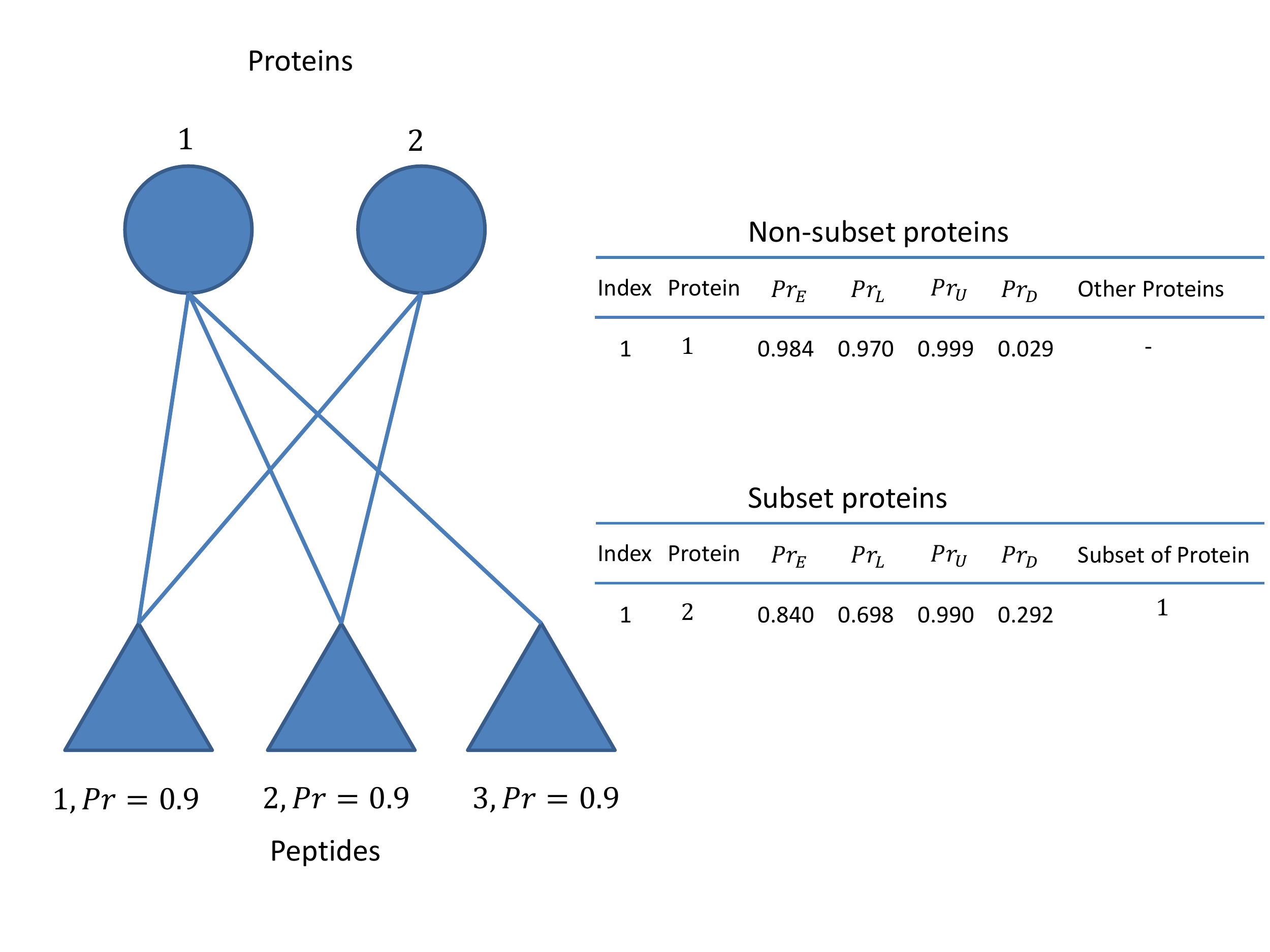}}
\end{figure}

An example is shown in Figure \ref{fig:example} for the illustration purpose. Protein 1 is more confident than protein 2. From data explanation's viewpoint, protein 1 is present whereas protein 2 is absent. This is because including protein 2 in the final protein list will not improve the data explanation efficiency. When we know that homogeneous proteins are present and desire more proteins, we can merge subset and non-subset proteins to obtain the final result by filtering proteins with thresholds on $\tpr_E$ and $\tpr_D$.

In our experiment, only non-subset proteins are considered in the comparison study.

\subsection{Protein Identification Results on Four Datasets}
\begin{figure}[H]
\caption{Protein identification results on four datasets. In (a) and (b), the curves of false versus true for the ISB dataset and the Sigma49 dataset are plotted; in (c) and (d), we show the curves of decoy versus target for the ISB dataset and the Sigma49 dataset; in (e) and (f), the curves of decoy versus target for the Human dataset and the Yeast dataset are shown. Considering that the performances measured by different validation methods may differ from each other, we also draw the curves of decoy versus target for the ISB dataset and the Sigma49 dataset.\label{fig:idresult}}
\centerline{\includegraphics[width=\linewidth]{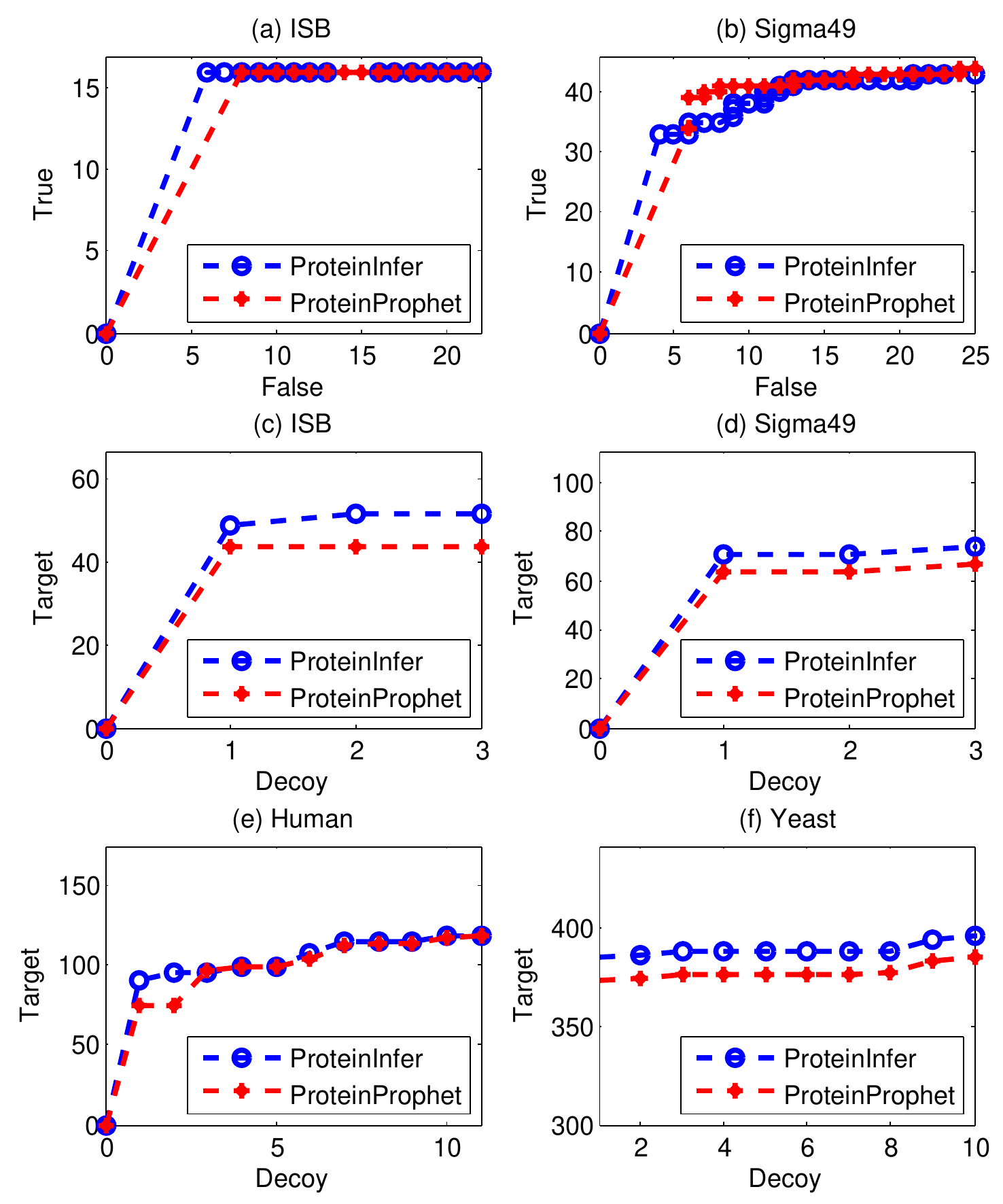}}
\end{figure}
The protein identification results on four datasets are shown in Figure \ref{fig:idresult}. Our method outperforms ProteinProphet in Figure \ref{fig:idresult}(a) and achieves a comparable result with ProteinProphet in Figure \ref{fig:idresult}(b). When using the curve of the decoy number versus the target number, our method dominantly outperforms ProteinProphet in Figure \ref{fig:idresult}(c)-(f).

\begin{table}[H]
\caption{Running time of our program compared with ProteinProphet. The total running time includes loading the peptide identification result, estimating protein probabilities and reporting the final result.\label{tab:time}}
\centering
\begin{tabular}{ccccc}
\hline
Program&ISB&Sigma49&Human&Yeast\\
\hline
ProteinInfer&0.340s&0.478s&1.057s&0.920s\\
ProteinProphet&14.273s&15.103s&16.473s&14.710s\\
\hline
\end{tabular}
\end{table}

All formulations of our method are closed-form. Thus, our method can calculate protein probabilities efficiently. In Table \ref{tab:time}, we show the running time of our method compared with ProteinProphet. The comparison is conducted on a computer with 4GB memory and Intel(R) Core(TM) i5-2500 CPU running the 32bit Windows 7 operating system. The final result shown in the table is the average running time of ten runs. The total running time includes loading the peptide identification result, estimating protein probabilities and reporting the result. The comparison of running time indicates the efficiency of our method in calculating protein probabilities.

According to the experimental results on four public available datasets, our method achieves competitive performance with ProteinProphet in a more efficient manner.

\subsection{The Parameter Issue}
In the preprocessing step, there are two parameters $\lambda_1$ and $\lambda_2$ corresponding to the expected number of unique peptides of true proteins and false proteins, respectively. These two parameters are estimated empirically from data. Alternatively, these two parameters can be set manually. Here, the performances of our method with different parameter settings are compared to show whether our method is sensitive to the parameter setting.

In this section, we conduct our experiment on the ISB dataset and the Sigma49 dataset. These two datasets have groundtruth, which reflects the impacts of parameter settings accurately.

\begin{table}[H]
\caption{Parameter settings. The default empirical parameter setting is marked with ``*''. In the experiment on each dataset, the performance based on the default parameter setting is shown for reference.\label{tab:parameter}}
\centering
\begin{tabular}{ccc}
\hline
Index&ISB&Sigma49\\
\hline
1&\tabincell{c}{$\lambda_1=5$\\$\lambda_2=1$}&\tabincell{c}{$\lambda_1=2$\\$\lambda_2=1$}\\
\hline
2&\tabincell{c}{$\lambda_1=15$\\$\lambda_2=1$}&\tabincell{c}{$\lambda_1=9$\\$\lambda_2=1$}\\
\hline
3&\tabincell{c}{$\lambda_1=10$\\$\lambda_2=1$}&\tabincell{c}{$\lambda_1=5$\\$\lambda_2=1$}\\
\hline
4&\tabincell{c}{$\lambda_1=12$\\$\lambda_2=5$}&\tabincell{c}{$\lambda_1=7$\\$\lambda_2=3$}\\
\hline
5&\tabincell{c}{$\lambda_1=12$\\$\lambda_2=10$}&\tabincell{c}{$\lambda_1=7$\\$\lambda_2=5$}\\
\hline
\end{tabular}
\end{table}

\begin{figure}[H]
\caption{The performances of our method on the ISB dataset and the Sigma49 dataset under different parameter settings shown in Table \ref{tab:parameter}.\label{fig:issigmabpara}}
\centerline{\includegraphics[width=\linewidth]{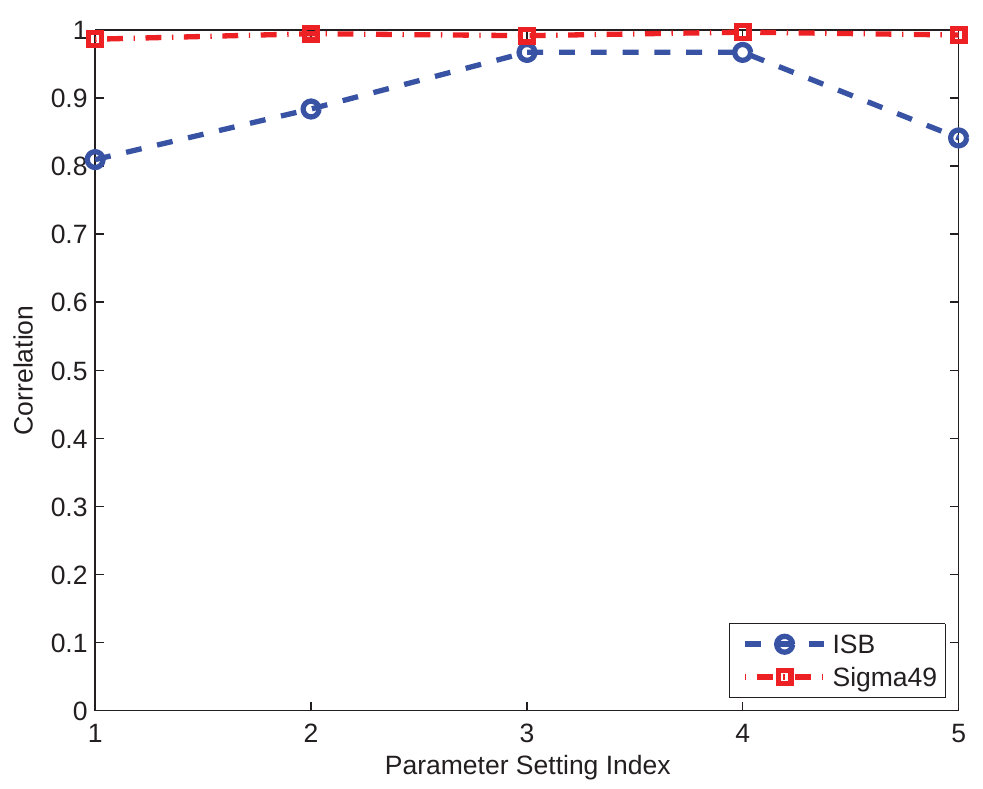}}
\end{figure}

The empirical estimations of $\lambda_1$ for the ISB dataset and the Sigma49 dataset are 12 and 7, respectively. The parameter settings we consider are shown in Table \ref{tab:parameter}. Under each parameter setting, we use the curve of false positives versus true positives to measure the corresponding performance of our method. The curve obtained by using the empirical parameters is taken as a reference. We calculate the correlation of other curves with the reference to illustrate the performance variation in different conditions. Figure \ref{fig:issigmabpara} shows the results.

\begin{figure}[H]
\caption{The performances of our method when parameters are set as $\lambda_1<\lambda_2$.\label{fig:misspec}}
\centerline{\includegraphics[width=\linewidth]{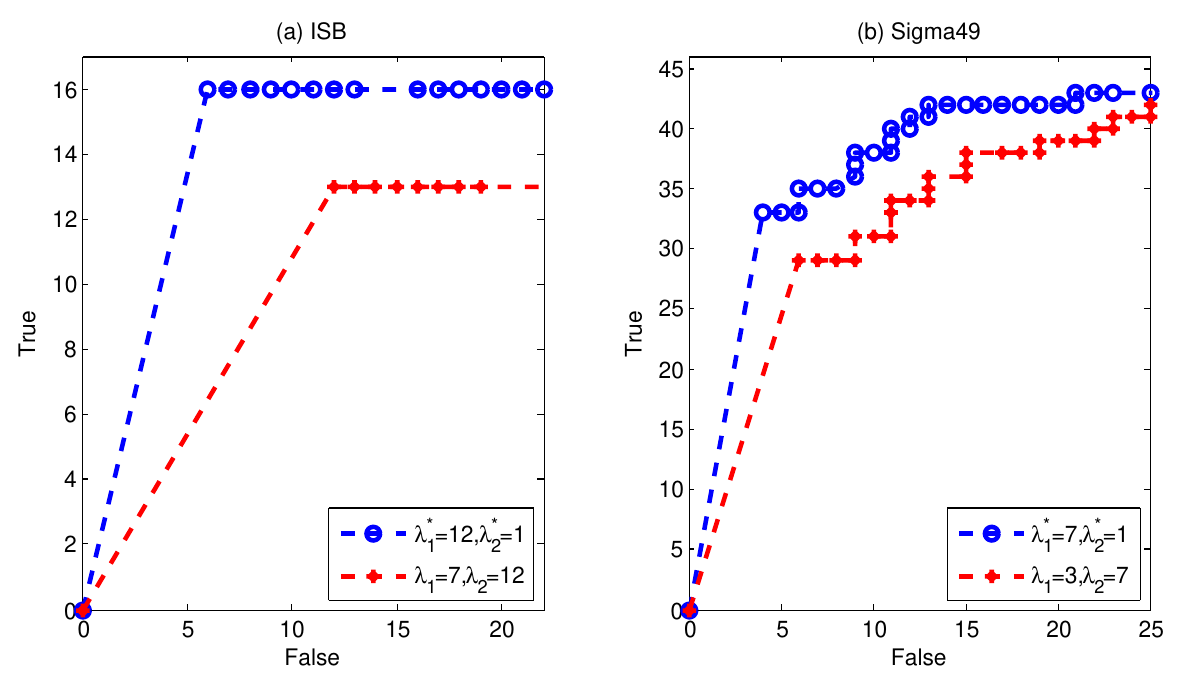}}
\end{figure}

In our model, we require that $\lambda_1>\lambda_2$. We also conduct experiments to show what the result is when parameters are misspecified (i.e. $\lambda_1<\lambda_2$). The result is shown in Figure \ref{fig:misspec}.

According to the results in two figures, we can see that our method is not sensitive to the parameter setting. The wrong parameter setting has a great impact on the protein identification result. Thus, we do not allow $\lambda_1 < \lambda_2$ in our program.

\subsection{More on Unique Peptide Probability Adjustment}
From the previous experiment, we can see that our method is not sensitive to the parameter setting. The only requirement is that parameters $\lambda_1>\lambda_2$. According to Figure \ref{fig:issigmabpara}, the performances under different parameters are close. Readers may be interested to know why we should adjust peptide probabilities.

\begin{table}[H]
\caption{The protein probabilities of decoy proteins before and after peptide probability adjustment. The last column is the rank position of the protein in the corresponding result. In the table, $\tpr_D$ are 0.0000 because peptides of decoy proteins tend to be unique. According to our inequality (18), $\tpr_L=\tpr_U$ when all peptides are unique.\label{tab:decoyprob}}
\centering
\begin{tabular}{cccccc}
\hline
\multicolumn{6}{c}{Probabilities Without Adjustment}\\
\hline
Index&Dataset&Protein&$\tpr_E$&$\tpr_D$&Rank\\
1&ISB&decoy\_499748&0.9971&0.0000&46\\
2&ISB&decoy\_237394&0.9243&0.0000&54\\
3&ISB&decoy\_224201&0.8864&0.0000&55\\
4&Sigma49&decoy\_519997&0.9517&0.0000&62\\
5&Sigma49&decoy\_170817&0.9417&0.0000&64\\
6&Sigma49&decoy\_271930&0.8248&0.0000&74\\
\hline
\multicolumn{6}{c}{Probabilities With Adjustment}\\
\hline
Index&Dataset&Protein&$\tpr_E$&$\tpr_D$&Rank\\
1&ISB&decoy\_499748&0.0650&0.0000&50\\
2&ISB&decoy\_237394&0.0024&0.0000&54\\
3&ISB&decoy\_224201&0.0016&0.0000&55\\
4&Sigma49&decoy\_519997&0.2548&0.0000&72\\
5&Sigma49&decoy\_170817&0.2190&0.0000&73\\
6&Sigma49&decoy\_271930&0.0755&0.0000&77\\
\hline
\end{tabular}
\end{table}

Let us consider the top three decoys proteins in the experiments on the ISB dataset and the Sigma49 dataset for the illustration purpose. Table \ref{tab:decoyprob} shows the protein probabilities of decoy proteins before and after peptide probability adjustment. According to the result, the protein probabilities of decoy proteins are decreased and they are ordered behind more target proteins. Decoy proteins are representative of a kind of error in protein identification. It is not desired to detect a decoy protein with a high confidence (e.g. decoy\_499748 is detected with $\tpr_E=0.9971$ and $\tpr_D=0.0000$). In this sense, the protein identification result becomes more meaningful after unique peptide adjustment. High confident decoy proteins are detected because its corresponding decoy peptides are detected with high confidence. Since peptide probability calculation is not perfect, we need to adjust it to achieve a more meaningful protein identification result.

In conclusion, unique peptide probability adjustment can improve the protein identification result (i.e. decoys proteins are ranked behind more target proteins) and make the result more meaningful than the result before adjustment. Thus, keeping the adjustment procedure in our program is essential.

\section{Discussions and Conclusions}
\subsection{Protein Probability Interval}
Protein probability interval $\tpr_D$ can be used to improve the distinction of protein identification results as well as to filter protein identification results.

\begin{table}[H]
\caption{The number of indistinguishable proteins based on probabilities without and with $\tpr_D$.\label{tab:distinct}}
\centering
\begin{tabular}{ccc}
\hline
Dataset&Without $\tpr_D$&With $\tpr_D$\\
\hline
ISB&27&21\\
Sigma49&38&36\\
Human&60&58\\
Yeast&181&178\\
\hline
\end{tabular}
\end{table}

Table \ref{tab:distinct} shows the numbers of indistinguishable proteins (based on the protein probability) without and with $\tpr_D$ on four datasets. From the result, we can see that $\tpr_D$ decreases the number of indistinguishable proteins.

\begin{table}[H]
\caption{The number of subset proteins without and with $\tpr_D$. In the table, ``$\&\&$'' means logical ``AND''.\label{tab:numsubset}}
\centering
\begin{tabular}{ccc}
\hline
Dataset&$\tpr_E\geq 0.9$&$\tpr_E\geq 0.9 \&\&\tpr_D\leq 0.02$\\
\hline
ISB&325&15\\
Sigma49&110&4\\
Human&12&4\\
Yeast&126&0\\
\hline
\end{tabular}
\end{table}

More importantly, $\tpr_D$ can be used as an extra filtering standard when $\tpr_E$ alone does not work effectively. This is very useful in the case when subset proteins are considered (e.g. protein identification rate is not satisfactory and homogeneous proteins are known to be present). Table \ref{tab:numsubset} shows the number of subset proteins when using $\tpr_E\geq 0.9$ and $\tpr_E\geq 0.9 \&\&\tpr_D\leq 0.02$ as filters, respectively. Here, ``$\&\&$'' means logical ``AND''. A great number of subset proteins can be filtered out by using $\tpr_D$. Thus, $\tpr_D$ and $\tpr_E$ form an effective filter to pick confident proteins.

\subsection{More on the Distinction of Protein Identification Results}
\begin{figure}[H]
\caption{The performance of ``ProteinInfer+Unique'' is obtained by considering the number of unique peptides as extra information in determining the order of proteins.\label{fig:sortunique}}
\centerline{\includegraphics[width=\linewidth]{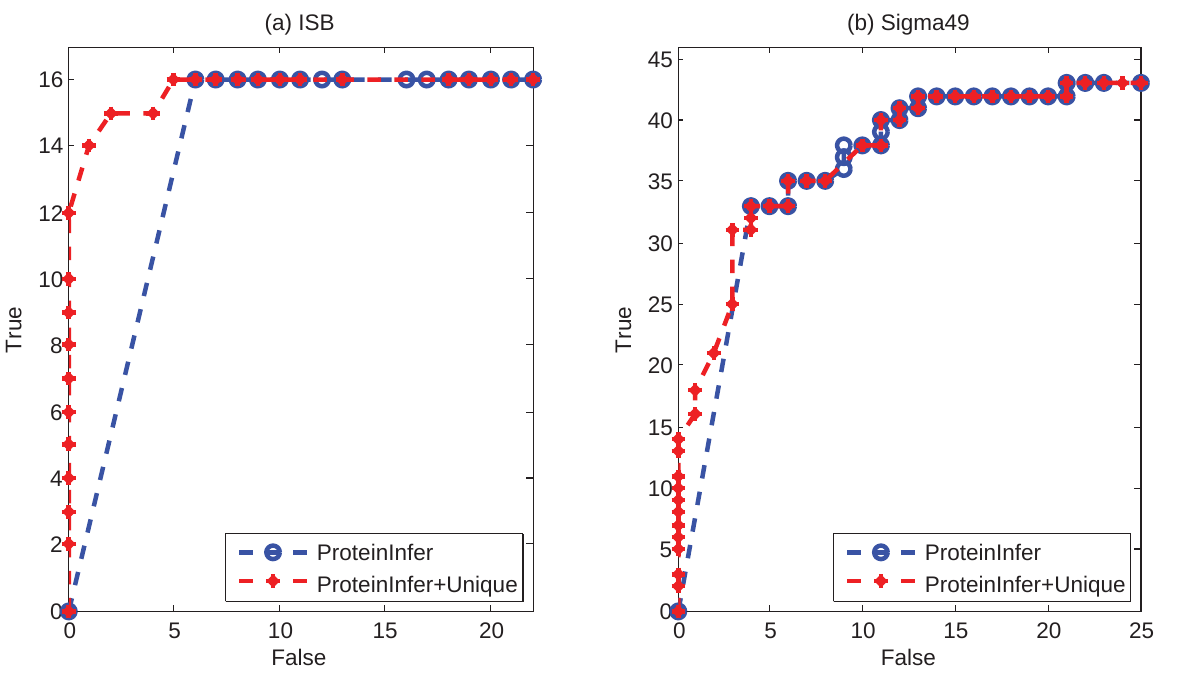}}
\end{figure}

When calculating protein probabilities, we often find that many proteins are assigned with the maximal score of one. The phenomenon can be explained with our model by considering the following example:
\begin{itemize}
\item Suppose a protein has three unique peptides with probabilities 0.97.
\item The protein probability based on the three unique peptides is $1-(1-0.97)^3=0.999973$.
\item Unique peptides are important to protein inference. According to equation (21), any extra identified peptides will further increase the confidence of the protein. Thus, we have $0.999973\leq \tpr_E\leq 1.0$ and $0\leq \tpr_D\leq 0.000027$. When only four decimal places are shown, we will have $\tpr_E=1.0000$ and $\tpr_D=0.0000$. Actually, many proteins are assigned to probability 1.0 because small numeric errors are ignored.
\end{itemize}
The poor distinction is mainly caused by the ignorable numeric errors. Considering the importance of unique peptides in protein inference, the distinction can be improved by sorting proteins with score one according to the number of unique peptides in descending order. The result is shown in Figure \ref{fig:sortunique}. In the figure, the performance of ``ProteinInfer+Unique'' is obtained by considering the number of unique peptides as an extra information in determining the order of proteins. This trick can be used when reporting the protein identification result.

However, this strategy does not work when two confident proteins have the same number of unique peptides. The key factor in estimating distinct protein probabilities is the peptide probability calculation. Unique peptides are important to protein inference and degenerate peptides will further increase the confidence of a protein. If the probability of a protein computed from its unique peptide is high, the protein must be confident. Thus, we need to estimate peptide probabilities conservatively especially for unique peptides.

\subsection{Conclusion}
In this paper, we propose a combinatorial perspective of the protein inference problem. From this perspective, we obtain the closed-form formulations of the lower bound, the upper bound and the empirical estimations of protein probability. Based on our model, we study an intrinsic property of protein inference: unique peptides are important to the protein inference problem and the impact of a degenerate peptide is determined by the number of times that the peptide is shared. In our experiments, we show that our concise model achieves competitive results with ProteinProphet.

\begin{acknowledgement}
This work was supported by the research proposal competition award RPC10EG04 from The Hong Kong University of Science and Technology, and the Natural Science Foundation of China under Grant No. 61003176.
\end{acknowledgement}

\bibliographystyle{achemso}
\bibliography{proteinid}

\end{document}